\documentclass[submission,copyright,creativecommons]{eptcs}
\providecommand{\event}{MeTRiD 2018} 
\usepackage{breakurl}             
\usepackage{underscore}           
\usepackage{wrapfig}
\usepackage{subcaption}
\usepackage{hyperref}

\usepackage{array}              
\usepackage{colortbl}           
\usepackage{booktabs}           

\usepackage{textcomp}           

\newlength{\tabwidthi}
\newlength{\tabwidthii}

\newcommand{\tabsize}{\footnotesize}

\usepackage{graphicx}
\graphicspath{{figures/}}

\usepackage{soul}
\soulregister\cite7
\soulregister\citep7
\soulregister\secn7
\soulregister\ref7
\soulregister\pageref7
\soulregister{\emph}{7}
\soulregister{\em}{0}
\soulregister{\textbf}{7}
\soulregister{\bf}{0}
\soulregister{\fig}{7}
\soulregister{\ie}{0}
\soulregister{\etc}{0}
\soulregister{\eg}{0}
\soulregister{\wrt}{0}
\soulregister{\resp}{0}
\soulregister{\cf}{0}
\soulregister{\mdash}{0}

\usepackage[colorinlistoftodos,bordercolor=white]{todonotes}

\newcommand{\fig}[1]{Figure~\ref{fig:#1}}
\newcommand{\Fig}[1]{Figure~\ref{fig:#1}}
\newcommand{\secn}[1]{Section~\ref{secn:#1}}
\newcommand{\Secn}[1]{Section~\ref{secn:#1}}
\newcommand{\tab}[1]{Table~\ref{tab:#1}}
\newcommand{\Tab}[1]{Table~\ref{tab:#1}}

\newcommand{\mdash}[1][~]{#1---#1}
\newcommand{\ie}[1][\ ]{i.e.#1}
\newcommand{\etc}[1][\ ]{etc.#1}
\newcommand{\eg}[1][\ ]{e.g.#1}
\newcommand{\cf}[1][\ ]{cf.#1}
\newcommand{\wrt}[1][\ ]{w.r.t.#1}
\newcommand{\resp}[1][\ ]{resp.#1}

\title{System Design in the Era of IoT\mdash Meeting the Autonomy Challenge\\(Invited paper)}

\author{Joseph Sifakis
\institute{University Grenoble Alpes and CNRS / Verimag\\ Grenoble, France}
\email{Joseph.Sifakis@univ-grenoble-alpes.fr}
}

\def\titlerunning{System Design in the Era of IoT}
\def\authorrunning{J. Sifakis}
\begin{document}
\maketitle

\begin{abstract}
The advent of IoT is a great opportunity to reinvigorate Computing by
focusing on autonomous system design.  This certainly raises
technology questions but, more importantly, it requires building new
foundation that will systematically integrate the innovative results
needed to face increasing environment and mission complexity.

A key idea is to compensate the lack of human intervention by adaptive
control.  This is instrumental for system resilience: it allows both
coping with uncertainty and managing mixed criticality services.  Our
proposal for knowledge-based design seeks a compromise: preserving
rigorousness despite the fact that essential properties cannot be
guaranteed at design time.  It makes knowledge generation and
application a primary concern and aims to fully and seamlessly
incorporate the adaptive control paradigm in system architecture.
\end{abstract}


\section{Introduction}
\label{secn:1}


\subsection{The IoT vision}
\label{secn:1.1}

The IoT vision promises increasingly interconnected smart systems
providing autonomous services for the optimal management of resources
and enhanced quality of life.  These are used in smart grids, smart
transport systems, smart health care services, automated banking
services, smart factories, \etc[ ] Their coordination should be
achieved using a unified network infrastructure, in particular to
collect data and send them to the cloud which in return should provide
intelligent services and ensure global monitoring and control.

The IoT vision raises a lot of expectations and in our opinion, some
over-optimism about its short-term outcome and impact.  According to
analysts, the IoT consists of two segments of uneven difficulty.  One
segment is the Human IoT which will be a significant improvement of
Internet where the dominant type of interaction will be client-server:
increasingly intelligent services to satisfy semantically rich
requests.

The other segment is the so-called Industrial IoT which would
coordinate autonomous services and systems.  The big difference with
Human IoT is that the latter involves fast closed loops of cooperating
agents where human intervention is external to normal behavior.  For
instance, human operators might intervene to change parameters of
autonomous agents or to mitigate potentially dangerous situations.  It
goes without saying that autonomous agents in the Industrial IoT will
be critical as they will be the core of complex systems such as,
Intelligent Transport Systems, Smart Grids and other critical
infrastructure, e-health and financial services.

It is well understood that under the current state of the art the
Industrial IoT vision cannot be reached for several reasons.  The
first and main reason is \emph{poor trustworthiness} of
infrastructures and systems deployed over the Internet.  It is
practically impossible to guarantee safety and security of services
and systems for the simple reason that they have been built in an ad
hoc manner.  Safety and security are cross cutting issues.  They
differ from performance in that they cannot be improved through
controlled experiments and tuning of system parameters.  A second
obstacle is the impossibility to guarantee reasonably short response
times in communications.  Most protocols in large scale systems are
not time-predictable.  Thus they cannot support reliable closed-loop
interaction of autonomous agents.  Finally, the reliability of
rigorously designed critical systems may be impacted by flaws of large
systems to which they are connected.

It is important to note that over the past decades almost no progress
has been made to resolve these problems.  The Industrial Internet
Consortium\footnote{%
  \url{https://www.iiconsortium.org/}
} has been established since March 2014 \emph{``in order to accelerate
  market adoption and drive down the barriers to entry''}.  Despite
the fact that it gathers together hundreds of important industrial
partners, to the best of our knowledge it has not delivered any
results that could even slightly move the mentioned roadblocks.

It is also important to note that these obstacles and limitations do
not discourage key industrial players from developing ambitious
projects challenging the current state of the art.  This is
particularly visible in automotive industry where the business stakes
and technological risks are huge.  In this race for increasingly
autonomous cars, the temptation is big to ``jump ahead'' and disrupt
to a large extent the standard rigorous systems engineering practice.
A typical example is customization by software updates on a monthly
basis for Tesla cars.  Such a practice breaches the rules of critical
systems standards which do not allow any modification of a product
after commercialization.  These standards require that the
trustworthiness of a product be fully established at design time.
Typically, an aircraft is certified as a product that cannot be
modified including all its HW components\mdash aircraft makers
purchase and store an advance supply of the microprocessors that will
run the software, sufficient to last for the estimated 50 year
production!

There is currently a striking contrast between the ambition for
increasingly large autonomous systems in the framework of IoT and the
lack of adequate rigorous design methods and supported by tools.  This
makes impossible the application of the current safety standards which
as a rule require \emph{conclusive evidence} that the built system can
cope with any type of critical mishap.


\subsection{What happened to the promise of rigorous, disciplined
  systems engineering?}
\label{secn:1.2}

The fulfilment of the vision for increasingly autonomous integrated
systems is not only a matter of maturity of the state of the art.  It
also depends on the degree of risks that society accepts to take in
exchange of considerable anticipated benefits.  The old ambition that
computing systems engineering should be as predictable as civil or
electrical engineering has drastically evolved within the Computing
community and the broader public.

We believe that this observed shift of opinion is largely due to the
lack of relevant theory enabling rigorous and disciplined design.
Theory has had a decreasing impact on software and systems engineering
over the past decades.  Formal methods failed to deliver results that
would raise computing systems engineering to the status of mature
engineering disciplines.  It turned out that such an ambition is
totally wrong and misleading.  Physical systems engineering relies on
theory allowing predictability and constructivity; and there are good
reasons to believe that no such a ``nice theory'' could exist for
computing systems.

A very common attitude of researchers is to work on mathematically
clean theoretical frameworks no matter how relevant they can be.
Mathematical clarity and beauty attracts the most brilliant who
develop ``low-level theory'' that often has no point of contact with
real practice.  The results are usually structure-agnostic and cannot
be applied to real-languages, architectures and complex systems built
from components and protocols.

Contrary to theoretical, practically-oriented research has developed
frameworks for programming, modeling and building complex systems in
an ad hoc manner\mdash involving a large number of constructs and
primitives, with little concern about rigorousness and semantics
(minimality, expressiveness).  These frameworks are badly amenable to
formalization.

The gap between theory and practice in systems engineering and the
huge push for system integration fueled by market needs and by
aspiration for innovation, have resulted in a dramatic change of the
opinion of IT professionals and by extension of the public opinion.
This is characterized by the following three synergizing positions.

One expresses a kind of ``resigned realism'' by adopting the idea that
we have to move forward and accept the risks because the benefits
resulting from the increasing automation of services and systems would
be much larger.  In an article by Vinton Cerf with the eloquent title
``Take two aspirin and call me in the morning'' \cite{ref1} one can
read: \emph{``So where does this leave us? I am fascinated by the
  metaphor of cyber security as a public health problem.  Our machines
  are infected and they are sometimes also contagious.  Our reactions
  in the public health world involve inoculation and quarantine and we
  tolerate this because we recognize our health is at risk if other
  members of society fail to protect themselves from infection.''}
Clearly this ``cyber-hygiene'' metaphor suggests that the current
situation is a fatality we cannot escape.  It gets us very far from
the vision of the engineer who designs buildings that will not
collapse with a very high probability for centuries.  More recently,
Warren Buffet talking about cyber insurance, has warned
that~\cite{ref2} \emph{``there's about a 2\% risk of a \$400 billion
  disaster occurring as a result of a cyber-attack or of other
  issue''}.  Buffett also explains that when he speaks to
cybersecurity experts, they tell him that \emph{``the offense is
  always ahead of the defense, and that will continue to be the
  case.''}  And he adds that \emph{``After all, the world runs on
  software, and software is written by humans who are just as flawed
  as you and me.''}  No doubt, such statements open the way for
accepting the imponderable risks induced by the ubiquitous and
extensive use of poorly engineered systems and applications of
unmanaged complexity.

A second position consists in showing a non-justified over-optimism
claiming that things will improve just by magic without drastically
changing the way we design systems and the network infrastructure.
The quote below is from a technology analyst at Davos WEF 2016:
\emph{``\textbf{There is no such thing as a secure system,} [\dots] As
  we give access to devices around us, from drones to thermostats, we
  need to make sure they cannot be easily hijacked.  \textbf{There
    will be a learning curve before we make them robust, but we'll
    learn}.''}  Similar opinions can be found in many articles
discussing the future of IoT in both the technical and broad public
press.

Furthermore, fallacious arguments about AI come to the aid of
over-optimists: \emph{``I really consider autonomous driving a solved
  problem. [\dots] I think we are probably less than two years
  away.''}\mdash Elon Musk, June 2, 2016.  Although AI will be key for
achieving autonomy, it does not help with making system design as
flawless as possible.

A third increasingly widespread opinion openly questions the interest
of theoretical foundations for software and system design.
Large-scale systems developers (\eg web-based systems) privilege
purely empirical approaches.

In an article published in CACM and entitled ``A new software
engineering'' one can read amongst others \cite{ref3}: \emph{``One
  might suggest computer science provides the underlying theory for
  software engineering\mdash and this was, perhaps, the original
  expectation when software engineering was first conceived.  In
  reality, however, computer science has remained a largely academic
  discipline, focused on the science of computing in general but
  mostly separated from the creation of software-engineering methods
  in industry.  While ``formal methods'' from computer science provide
  the promise of doing some useful theoretical analysis of software,
  practitioners have largely shunned such methods (except in a few
  specialized areas such as methods for precise numerical
  computation).''}  Such positions, especially when published in a
flagship ACM journal, are likely to have a deep and irreversible
impact.  They strikingly contrast with the Formal Methods vision
advocated forty years ago by pioneers of Computing.


\subsection{The Way Forward}
\label{secn:1.3}

The purpose of this paper is to discuss to what extent the IoT vision
is reachable under the current state of the art.  Starting from the
premise that current critical system practice and standards are not
applicable to autonomous systems in the context of IoT, we identify
the new factors of difficulty/complexity and propose novel avenues for
overcoming them.

It is well-understood that systems engineering comes to a turning
point moving: 1)~from small size centralized non evolvable systems, to
large distributed evolvable systems; 2)~from strictly controlled
system interaction with its external environment, to non-predictable
dynamically changing environments; 3)~from correctness at design time
to correctness ensured through adaptation.

It is urgent that research in Computing refocuses on the so many open
problems raised by modern system design, breaking with the
``positivist'' spirit of Formal Methods and working on real problems
maybe at the detriment of ``theoretical purity''.  This is the only
way to refute statements such as \emph{``system design is a definitely
a-scientific activity driven by predominant subjective factors that
preclude rational treatment''}.  Similar positions are promoted by
influential ``guilds'' of gurus, craftsmen and experts within big SW
companies as well as by the whole ecosystem of technology consulting
companies.  The development of a booming market in cybersecurity, and
vested interests of consulting companies are hindering public
awareness for more trustworthy systems and infrastructure.

We need to reassess existing methods in the light of the needs as they
have changed over the past decades.  Clearly, verification and formal
methods should be applied to small systems whenever it is realistic
(cost-effective and tractable).  We should investigate alternative
methods for achieving correctness not suffering complexity limitations,
\eg by construction.

We should admit that in the context of IoT, even critical systems
cannot be guaranteed exempt of flaws at design time.  Without giving
up the requirement for rigorousness, we should seek tradeoffs for
deciding if a system is trustworthy enough for the intended use.

The focus should be on autonomous system design, addressing related
specific needs.  It is essential to study the concept of autonomy and
identify the key theoretical and technical results for taking up the
autonomy challenge.

Autonomy is understood as the capacity of an agent (service or system)
to achieve a set of coordinated goals by its own means (without human
intervention) by adapting to environment variations.  It covers three
different aspects: 1)~autonomy of decisions \ie choosing among
possible goals; 2)~autonomy of operations planned to achieve the
goals; 3)~autonomy of adaptation \eg by learning.

The degree of autonomy of a system can be captured as the product of
three independent factors: 1)~Complexity of the environment;
2)~complexity of mission and its implementation as a sequence of
feasible tasks; 3)~non-intervention of human operators.

\begin{table}[ht!]
  \tabsize
  \centering
  \caption{SAE vehicle autonomy levels \cite{ref15,ref5}}
  \label{tab:1}
  \settowidth{\tabwidthi}{\bf Level 0}
  \setlength{\tabwidthii}{\textwidth-\tabwidthi-2\tabcolsep}
  \begin{tabular}{@{}
      >{\bf}l
      >{\raggedright}p{\tabwidthii}    
      @{}
    }
    \toprule
    Level 0 &
    \textbf{No automation}
    \tabularnewline\midrule
    Level 1 &
    \textbf{Driver assistance required (``hands on''):}
    \tabularnewline
    & The driver still needs to maintain full situational awareness
    and control of the vehicle, \eg cruise control.
    \tabularnewline\midrule
    Level 2 &
    \textbf{Partial automation options available(``hands off''):}
    \tabularnewline
    &
    Autopilot manages both speed and steering under certain
    conditions, \eg highway driving.
    \tabularnewline\midrule
    Level 3 &
    \textbf{Conditional Automation(``eyes off''):}
    \tabularnewline
    &
    The car, rather than the driver, takes over actively monitoring
    the environment when the system is engaged.  However, human drivers
    must be prepared to respond to a ``request to intervene''.
    \tabularnewline\midrule
    Level 4 &
    \textbf{High automation (``mind off''):}
    \tabularnewline
    &
    Self driving is supported only in limited areas (geofenced) or
    under special circumstances, like traffic jams.
    \tabularnewline\midrule
    Level 5 &
    \textbf{Full automation (``steering wheel optional''):}
    \tabularnewline
    &
    No human intervention is required, \eg a robotic taxi.
    \tabularnewline\bottomrule
  \end{tabular}
\end{table}

The interplay between these three factors is illustrated by the six
SAE autonomy levels shown in \tab{1} varying from Level~0, for no
automation, to Level~5, for full automation.  Level~4 brings
restrictions to the environment as self-driving is supported only in
limited areas or under special circumstances, like traffic jams.
Level~3 is a critical level as the human driver must be prepared to
respond to a ``request to intervene''.  This type of interaction with
a passive driver suddenly solicited to take over, raises some issues
as attested by the recent accident of an Uber car \cite{ref4}.

In this paper, we propose research directions for each one of the
three factors characterizing the degree of system autonomy:

\begin{itemize}
\item To cope with environment complexity we should seek a tighter
  integration of computing elements and their environment, in
  particular through the use of \emph{cyber-physical components}.

\item To cope with mission complexity we need dynamic reconfiguration
  of resources and self-or\-ga\-ni\-za\-tion in particular through the use of
  adequate \emph{architectures};

\item To compensate the lack of direct human intervention, we advocate
  extensive use of \emph{adaptive control techniques}.
\end{itemize}

The paper is structured as follows.  \Secn{2} summarizes well-known
facts about system design.  It discusses current limitations of
critical system design as it is enforced by standards, and its failure
to satisfy current needs.  \Secn{3} presents research avenues for
coping with the complexity stemming from the need for increasingly
high autonomy.  \Secn{4} advocates Knowledge-Based Design as an
alternative to existing critical system design techniques.  \Secn{5}
concludes with a discussion about scientific, technological and
societal stakes of the autonomy challenge.


\section{System Design}
\label{secn:2}


\subsection{About system correctness}
\label{secn:2.1}

System correctness is characterized as the conjunction of two types of
properties: trustworthiness and optimization properties \cite{ref6}.
Trustworthiness means that the system can be trusted, and that it will
behave as expected despite: 1)~software design and implementation
errors; 2)~failures of the execution infrastructure; 3)~interaction
with potential users including erroneous actions and threats; and
4)~interaction with the physical environment including disturbances
and unpredictable events.  Optimization requirements demand the
optimization of functions subject to constraints on resources such as
time, memory, and energy, dealing with performance and
cost-effectiveness.

System designers should ensure trustworthiness without disregarding
optimization as the two types of requirements are often antagonistic.
For small size critical systems, emphasis is put on trustworthiness,
while for large systems the emphasis is on optimization, provided that
system availability remains above some threshold.  Thus comes the
well-known concept of levels of criticality going from critical
systems to best-effort systems.  This distinction reflects a big
difference in development methods and costs.  Critical systems
development is subject to standards defining evaluation criteria
enforced by certification authorities.  According to standards
currently in effect, the reliability of an aircraft should be of the
order of $10^{-9}$ failures per hour.  Note that the reliability of
best-effort systems may be as low as $10^{-4}$ failures per hour while
the reliability of a rocket is around $10^{-6}$ failures per hour.
Existing empirical laws regarding the evolution of development costs
of high confidence systems show that multiplying the reliability of a
system by some factor may require exponentially higher development
costs.  Furthermore, the increase of the size of a system for the same
reliability level has a similar effect.

These facts explain the current gap between critical and best effort
system design discussed later.

To summarize, critical system design, as enforced by standards and
methodologies, costs a lot and is applicable only to small size
systems (\eg some hundreds of thousands of lines of code).


\subsection{Rigorous system design\mdash The principles}
\label{secn:2.2}

System design is the engineering process that leads from requirements
to a mixed HW/SW system meeting the requirements.  The process follows
a flow organized in steps, some of which are iterative.  At each step,
the designer enriches a model of the designed system assisted by tools
allowing him to check that the properties derived from the
requirements are met.

As achieving formal correctness seems practically impossible for
real-life systems, we have proposed rigorousness as a minimal
requirement for guaranteeing trustworthiness in system design
\cite{ref8, ref6}.  Rigorous system design is model-based and
accountable.

\paragraph{Model-Based}

\begin{figure}
  \centering
  \includegraphics[width=0.6\textwidth]{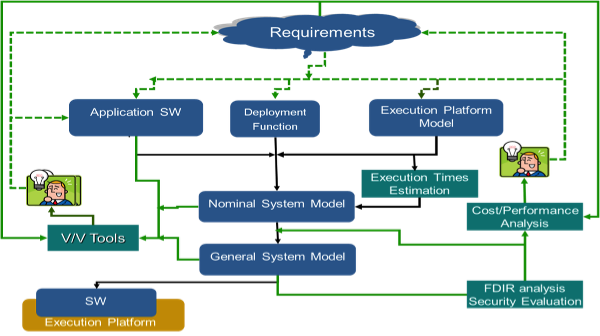}
  \caption{The principle of a rigorous system design flow}
  \label{fig:1}
\end{figure}

Even if different languages are used by designers, all these languages
should be embedded in a common \emph{host model} in order to guarantee
the overall coherency of the flow \cite{ref7} (\fig{1}).  In practice,
there is no need of a distinct semantic model.  The host model can be
software written in a general purpose programming language adequately
structured and annotated.

A key idea is that the application software model is ``composed'' with
an abstract execution platform model to get a \emph{nominal system
  model} that describes the behavior of the software running on the
platform.  The composition operation is specified through a deployment
consisting of 1)~a mapping assigning processes to processing elements
and data to memories of the platform; 2)~an associated scheduling
algorithm.

The nominal system model is progressively enriched by application of
property-preserving model-to-model transformations to obtain a
\emph{general system model}.  The transformations are local and do not
suffer any complexity limitations.  They should be proven correct in
the sense that they should preserve essential properties such as
invariants and deadlock-freedom.  They consist in adding, first,
timing information provided by an execution time estimation tool.
Then, other transformations add mechanisms for resilience to failures,
attacks or any kind of critical hazards.  The so enriched model
sufficiently validated is used to generate implementations.

\paragraph{Accountable}

Accountability concerns two aspects: 1)~evidence that the designer's
choices are justified by the need to meet properties implied by
requirements; 2)~guarantees that properties already established at
some design step, will still hold in subsequent steps.  In practice,
accountability can be eased through the use of assurance case
methodologies.  These allow structuring the designer's reasoning to
gain confidence that systems would work as expected \cite{ref9}.

Note that there exist only a few rigorous system design approaches
dedicated to the development of safety-critical real-time systems.
Some approaches are based on the synchronous programming paradigm
\cite{ref10}.  Others are represented mainly by flows based on the ADA
standard \cite{ref11}.  In both, rigorousness is achieved at the price
of severe restrictions regarding the programming and the execution
model \cite{ref6}.

We explain below why current rigorous system design methodologies
cannot meet the demand for increasing autonomy in the framework of
IoT.


\subsection{The two pillars of rigorous system design}
\label{secn:2.3}

\subsubsection{Verification}
\label{secn:2.3.1}

Verification is a process used to provide conclusive evidence that a
system is correct with respect to a given property.  This can be
achieved by checking in some exhaustive and rigorous manner that a
system model or the system development process model meets the
property.  Verification differs from testing in that it seeks
exhaustivity.  It requires special care in the elaboration of the
model and the formalization of the properties to be verified.  The
properties should be derived from requirements usually expressed in
natural language.  The models should be faithful: whatever property
holds for the real system should also hold for the model.

Verification and model checking in particular, constitutes one of the
main breakthroughs for quality assurance in both hardware and software
development.  It has drastically contributed to gaining mathematical
confidence in system correctness \cite{ref12}.  Nonetheless, a key
limitation of automated verification methods is that they are applied
to \emph{global} models.  Despite intensive research efforts, it was
not possible to develop automated compositional verification methods
(inferring global system properties from properties of the constituent
components at reasonable computational cost).  This is a serious
limitation of automated verification techniques.

\begin{figure}
  \centering
  \includegraphics[width=0.5\textwidth]{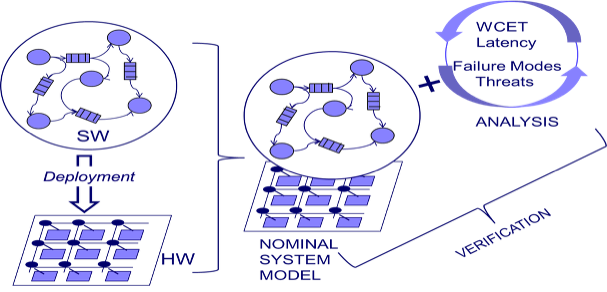}
  \caption{Building the general system model}
  \label{fig:2}
\end{figure}

Another limitation concerns the possibility of building faithful
general system models as defined in \secn{2.2}.  These models should
account for the system behavior in the presence of critical events
that may compromise correctness (\fig{2}).  Modeling the impact of
such events requires a very good knowledge of both the system and its
environment, in particular of the dynamics of their interaction.  As
system complexity and openness increase, the number of potentially
critical events explodes.  Note that the degree of difficulty is much
higher for security analysis.  While some methodologies exist for
fault detection, isolation and recovery (FDIR) analysis \cite{ref13}
there is no systematic approach to global security analysis as it is
hard to figure out how human ingenuity can exploit system
vulnerability.  The recently discovered Intel's security flaw is a
remarkable illustration of this fact \cite{ref14}.

The second roadblock to system verification is system openness
requiring the formalization of the interaction dynamics between the
system and some abstraction of its environment.  It is very hard to
formalize requirements involving human behavior, computers and
electromechanical devices.

\Tab{2} provides a list of high-level requirements (behavioral
competencies) for self-driving cars proposed by the California PATH
project \cite{ref15}.  Their decomposition into formalized properties
to be checked on autonomous driving systems seems to be an
unsurmountable problem.  To formally verify each one of these
requirements, it is necessary to decompose it into a set of properties
involving physical quantities, discrete system events as well as
relevant information about the geometry of the external environment,
\eg coming from analysis of sensory information.

All the above limitations are even worsened by the fact that modern
machine learning techniques, essential components of autonomous
systems, cannot be verified.  The simple and obvious reason is that
they are not developed based on requirements, \eg  that specify how a
dog looks different from a cat.  They learn just like children learn
the differences between cats and dogs.  So, establishing their safety
according to existing standards is problematic.

Finally, we should not neglect some non-technical obstacles that have
to do with the social acceptance of the truthfulness of verification
processes: it is not sufficient to prove that a system is correct by
some possibly sophisticated method.  It is even much more important to
convince institutions, \eg certification authorities \cite{ref16}.
This requires special care in the development of verification
technology with the possibility to check that also the whole
verification process is exempt of error.

\begin{table}[ht!]
  \tabsize
  \centering
  \caption{Behavioral competencies for self-driving cars proposed by
    the California Path Project}
  \label{tab:2}
  \settowidth{\tabwidthi}{\bf 99.}
  \setlength{\tabwidthii}{0.9\textwidth-\tabwidthi-2\tabcolsep}
  \begin{tabular}{@{}
      >{\bf}l
      >{\raggedright}p{\tabwidthii}    
      @{}
    }
    \toprule
1.  & Detect and Respond to Speed Limit Changes and Speed Advisories
    \tabularnewline\midrule
2.  & Perform High-Speed Merge (\eg Freeway)
    \tabularnewline\midrule
3.  & Perform Low-Speed Merge
    \tabularnewline\midrule
4.  & Move Out of the Travel Lane and Park (\eg to the Shoulder for Minimal Risk)
    \tabularnewline\midrule
5.  & Detect and Respond to Encroaching Oncoming Vehicles
    \tabularnewline\midrule
6.  & Detect Passing and No Passing Zones and Perform Passing Maneuvers
    \tabularnewline\midrule
7.  & Perform Car Following (Including Stop and Go)
    \tabularnewline\midrule
8.  & Detect and Respond to Stopped Vehicles
    \tabularnewline\midrule
9.  & Detect and Respond to Lane Changes
    \tabularnewline\midrule
10. & Detect and Respond to Static Obstacles in the Path of the Vehicle
    \tabularnewline\midrule
11. & Detect Traffic Signals and Stop/Yield Signs
    \tabularnewline\midrule
12. & Respond to Traffic Signals and Stop/Yield Signs
    \tabularnewline\midrule
13. & Navigate Intersections and Perform Turns
    \tabularnewline\midrule
14. & Navigate Roundabouts
    \tabularnewline\midrule
15. & Navigate a Parking Lot and Locate Spaces
    \tabularnewline\midrule
16. & Detect and Respond to Access Restrictions (One-Way, No Turn, Ramps, etc.)
    \tabularnewline\midrule
17. & Detect and Respond to Work Zones and People Directing Traffic in Unplanned or Planned Events
    \tabularnewline\midrule
18. & Make Appropriate Right-of-Way Decisions
    \tabularnewline\midrule
19. & Follow Local and State Driving Laws
    \tabularnewline\midrule
20. & Follow Police/First Responder Controlling Traffic (Overriding or Acting as Traffic Control Device)
    \tabularnewline\midrule
21. & Follow Construction Zone Workers Controlling Traffic Patterns (Slow/Stop Sign Holders).
    \tabularnewline\midrule
22. & Respond to Citizens Directing Traffic After a Crash 
    \tabularnewline\midrule
23. & Detect and Respond to Temporary Traffic Control Devices
    \tabularnewline\midrule
24. & Detect and Respond to Emergency Vehicles
    \tabularnewline\midrule
25. & Yield for Law Enforcement, EMT, Fire, and Other Emergency Vehicles at Intersections, Junctions, and Other Traffic Controlled Situations
    \tabularnewline\midrule
26. & Yield to Pedestrians and Bicyclists at Intersections and Crosswalks
    \tabularnewline\midrule
27. & Provide Safe Distance From Vehicles, Pedestrians, Bicyclists on Side of the Road
    \tabularnewline\midrule
28. & Detect/Respond to Detours and/or Other Temporary Changes in Traffic Patterns
    \tabularnewline\bottomrule
  \end{tabular}
\end{table}

\subsubsection{The V-model}
\label{secn:2.3.2}

Systems engineering standards often recommend the so-called ``V
model'', which consists in decomposing system development into two
flows.  The first is top-down, starts from requirements and involves a
hierarchical decomposition of the system into components and a
coordinating architecture.  The other flow is bottom-up and consists
in progressively assembling, integrating, and testing the designed
components.  This model has been criticized for several reasons \cite{ref7}:

\begin{enumerate}
\item It assumes that all the system requirements are initially known,
  can be clearly formulated and understood.  Anyone with minimal
  experience in system design realizes that such an assumption is not
  realistic.  It is very often necessary to revise initial
  requirements.

\item It assumes that system development is top-down driven by
  refining the requirements and projecting them on components.  This
  assumption does not seem realistic too.  First, modern systems are
  never designed from scratch; they are often built by incrementally
  modifying existing systems and by extensive component reuse.
  Second, it considers that global system requirements can be broken
  down into properties satisfied by system components, which is a
  non-trivial problem.

\item It relies mainly on correctness-by-checking (verification or
  testing) which takes place bottom-up only after the implementation
  is completed.
\end{enumerate}

For all these reasons the V-model has been abandoned in modern
software engineering in favor of the so-called Agile methodologies.
These consider that coding and designing should go hand in hand:
designs should be modified to reflect adjustments made to the
requirements.  So, design ideas are shared and improved on during a
project ``in a spiral manner''.

In our opinion, the main merit of Agile methodologies is their
criticism of the V-model rather than a disciplined and well-structured
way for tackling system development.


\section{Trends and Challenges in Autonomous System Design}
\label{secn:3}

We have seen that system autonomy can be characterized as the
interplay between three complexity factors: environment, mission and
non-intervention of humans.  These factors do not impact autonomous
system design in the same manner.  While hardness increases for
increasing environment and mission complexity, increasing the degree
of automation may sometimes ease the design problem.

We discuss these three trends and show work directions and associated
challenges.


\subsection{System Design Complexity}
\label{secn:3.1}

\begin{figure}
  \centering
  \includegraphics[width=0.8\textwidth]{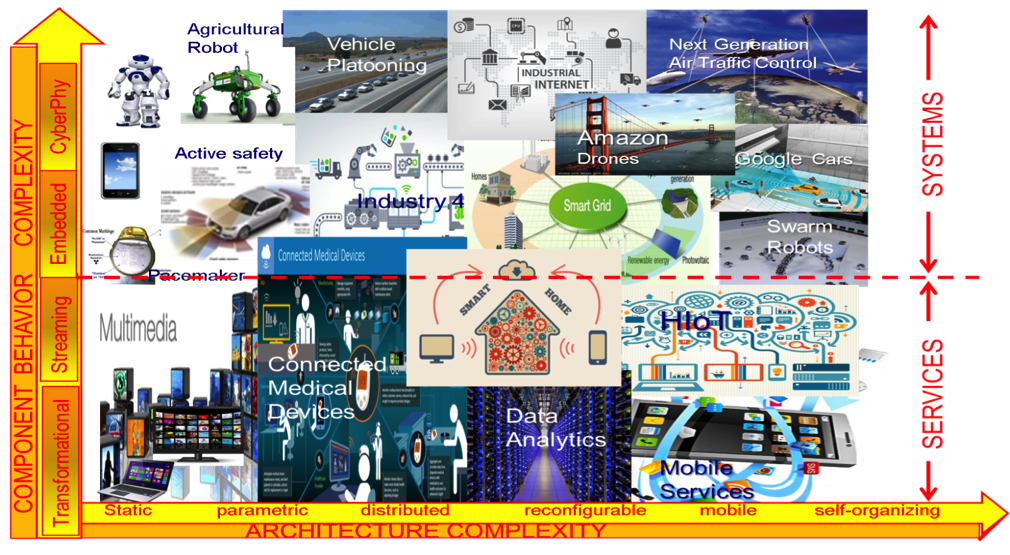}
  \caption{Autonomous system design complexity}
  \label{fig:3}
\end{figure}

We first analyze the role of the two factors directly impacting design
complexity.  Increasing autonomy requires tighter integration of
computers and their physical environment as well as self-organization
of system resources.  \Fig{3} illustrates this observation.  As we
move away from the origin, autonomy increases from purely functional
components to cyber physical components and from static to
self-organizing architectures.  The dashed line separating functional
and streaming components from embedded and cyber physical components,
marks the border between the Human and the Industrial IoT.  Increasing
architecture complexity reflects increasing autonomy for services and
systems respectively.

\subsubsection{Component Complexity\mdash Cyber Physical Systems}
\label{secn:3.1.1}

\begin{figure}
  \hfill
  \begin{minipage}{.27\textwidth}
    \centering
    \includegraphics{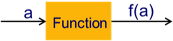}
    \subcaption{\tabsize Functions, methods,\\ \eg Client-Server systems}
    \label{fig:4:function}%

    \medskip
    \includegraphics{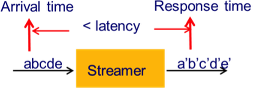}
    \subcaption{\tabsize Streaming systems,\\ \eg encoders, signal processing systems}
    \label{fig:4:streamer}%
  \end{minipage}
  \hfill
  \begin{minipage}{0.25\textwidth}
    \centering
    \includegraphics{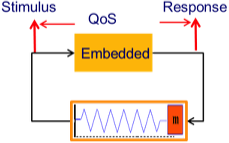}
    \subcaption{\tabsize Embedded systems,\\ \eg flight controller}
    \label{fig:4:embedded}%
  \end{minipage}
  \hfill
  \begin{minipage}{0.33\textwidth}
    \centering
    \includegraphics{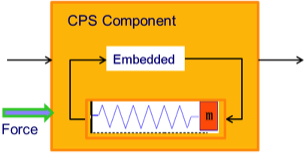}
    \subcaption{\tabsize Cyber-Physical Systems,\\ \eg self-driving cars}
    \label{fig:4:cps}%
  \end{minipage}
  \hspace*{\fill}
  \caption{Classification of system components}
  \label{fig:4}
\end{figure}

\Fig{4} illustrates different types of components for increasing
intricacy of interaction with their environment.  The simplest
components are functional. They compute some function $f$ by
delivering, for any input datum $x$ a corresponding output $f(x)$.
Streamers compute functions on streams.  For a given input stream of
values, they compute a corresponding output stream.  The output value
at some time $t$ depends on the history of the input value received by
$t$.  Encoders/decoders or in general data-flow systems involve
streamers.  The requirements for such components are functional
correctness and specific time-dependent properties such as latency.

Embedded components continuously interact with a physical environment
so as to ensure global properties.  Such components are mixed HW/SW
components, where the real-time behavior and dynamic properties are
essential for correctness.  Finally, cyber-physical components
integrate embedded and physical components.  They combine discrete and
continuous dynamics.  Cyber and physical aspects are deeply
intertwined and their composition requires multi-scale and
multi-domain integration of theories.

The study of cyber-physical systems has been the object of a rich
literature over the past decade.  Nonetheless, key problems raised by
their rigorous design remain open.  These have to do with the faithful
modeling of complex electromechanical systems with discrete events, as
well as the discretization of hybrid models in view of their
implementation.  We currently lack theory and supporting tools for
component-based modeling, as the concepts of composition of physical
and cyber models are radically different.  Physical systems models are
inherently declarative, synchronous, parallel and the interaction
between components is data-flow; on the contrary, computation is
inherently procedural, sequential and interaction is natively
event-driven.

Discretization of hybrid models raises semantic problems about how to
detect and precisely simulate converging system dynamics.  We also
lack theory for safe and efficient discretization as well as for
deciding whether a hybrid model is executable.  The interested reader
can find a detailed discussion of these issues in \cite{ref17}.

\subsubsection{Architecture complexity\mdash Self-organizing architectures}
\label{secn:3.1.2}

Architectures depict principles of coordination, paradigms that can be
understood by all, allow thinking on a higher plane and avoiding
low-level mistakes.  They are a means for ensuring correctness by
construction, as they enforce specific global properties
characterizing the coordination between the composed components.
System developers extensively use libraries of reference architectures
such as time-triggered architectures, security architectures and
fault-tolerant architectures.

Architectures can be considered as generic operators that can take as
arguments arbitrary numbers of instances of component types.  The
resulting system of coordinated components satisfies by construction a
characteristic property.  For instance, Client-Server architectures
are used to coordinate arbitrary numbers of instances of clients and
servers.  The ensured characteristic properties include atomicity of
transactions and fault-tolerance.

An architecture can be characterized by the three following
attributes:

\begin{enumerate}
\item The type of the coordinated components specified by their
  interfaces including port (function) names and rules regarding the
  way they are handled by the component's environment;

\item The type of the supported interactions, which may vary from
  point-to-point to multiparty interaction including rendezvous,
  broadcast, synchronous or asynchronous interaction;

\item The topology of the architecture which reflects the structure of
  the connections between components as well as between components and
  their environment.  Simple topologies include centralized
  architectures (components interacting through a single coordinator),
  hierarchical architectures, ring architectures and clique
  architectures.
\end{enumerate}

\begin{figure}
  \hfill
  \begin{minipage}{.2\textwidth}
    \centering
    \includegraphics{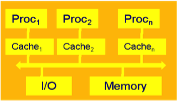}
    \subcaption{\tabsize Static architecture}
    \label{fig:5:static}%
  \end{minipage}
  \hfill
  \begin{minipage}{.23\textwidth}
    \centering
    \includegraphics{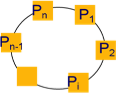}
    \subcaption{\tabsize Parametric architecture}
    \label{fig:5:parametric}%
  \end{minipage}
  \hfill
  \begin{minipage}{.27\textwidth}
    \centering
    \includegraphics{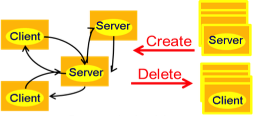}
    \subcaption{\tabsize Dynamic architecture}
    \label{fig:5:dynamic}%
  \end{minipage}
  \hspace{\fill}

  \hfill
  \begin{minipage}{.26\textwidth}
    \centering
    \includegraphics{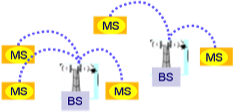}
    \subcaption{\tabsize Mobile architecture}
    \label{fig:5:mobile}%
  \end{minipage}
  \hfill
  \begin{minipage}{.36\textwidth}
    \centering
    \includegraphics{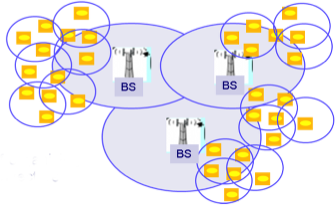}
    \subcaption{\tabsize Self-organising architecture}
    \label{fig:5:self-organising}%
  \end{minipage}
  \hspace{\fill}
  \caption{Classification of architectures}
  \label{fig:5}
\end{figure}

We characterize the complexity of architectures by the degree to which all these attributes may dynamically change and be organized, as illustrated in \fig{5}.  This classification distinguishes five types of architectures and has been carried out based on technical criteria presented in \cite{ref18,ref19}.

\begin{enumerate}
\item Static architectures involve a predefined number of components
  and interconnections, \eg HW architectures.

\item Parametric architectures take as arguments any number of
  instances of components of the appropriate type.  Protocols, SW
  architectures, distributed algorithms are parametric architectures.

\item Dynamic architectures are parametric architectures supporting
  component dynamism\mdash components may be created and deleted as in
  a Client-Server system.

\item Mobile architectures are dynamic architectures where
  additionally the system's external environment can dynamically
  change, as in mobile telecommunication systems.

\item Finally, self-organizing architectures allow reconfiguration
  between modes, where each mode has its own coordination rules.  Such
  architectures are instrumental for modeling complex autonomous
  systems, \eg swarms of robots and platooning vehicles.
\end{enumerate}

Although architectures are of paramount importance in systems
engineering, their formal study has not attracted so far the deserved
attention.  Hardware engineering relies on the concept of architecture
as a means of building systems that satisfy essential properties by
construction.  In software engineering the focus has mainly been on
the development of specific Architecture Description Languages.
Unfortunately, most of the effort deals with syntactic aspects.  More
than 100 such languages have been proposed over the past twenty years
but none of them has been adopted by practitioners
\cite{suggested-ref}.

We believe that research in architectures should focus on modeling
techniques for self-organizing architectures by studying basic
structuring principles and mechanisms and evaluating their
expressiveness.  Additionally, architectures should be used as a means
for building systems that are by construction correct.  The idea is
quite simple and straightforward.  Putting it into practice requires
work in the following two directions.

First, we should study architectures as parametric behavior
transformers and develop basic results for checking their correctness.
Existing results on parametric verification put emphasis on
limitations \cite{ref20,ref21}.  The parametric verification of very
simple parametric systems, even with finite-state components, is
intractable.  Nonetheless, we believe that it is worthwhile seeking
practically relevant results in this direction, \eg by developing
semi-decision methods as we did for infinite state systems.

Second, we need composability theory for architectures allowing to
combine two architectural solutions meeting each one a characteristic
property, into a single solution meeting the conjunction of the
properties \cite{ref22}.  Such results can be of tremendous practical
relevance.  Architectures can be implemented as software into which
components can be plugged.  Composability results simply guarantee the
interference-free composition of two architecture softwares preserving
respectively the enforced characteristic properties.


\subsection{Non-human intervention\mdash Adaptivity}
\label{secn:3.2}

When humans cooperate with semi-automated systems to achieve a
mission, \eg Level~2 and 3 in autonomous cars, they are mainly
responsible for handling uncertain situations.  It is well understood
that computers are not good enough when dealing with such situations,
while they perform better than humans specific well-defined tasks.  An
important question is how to design systems that exhibit adaptive
behavior in real-time exactly as humans do.

\subsubsection{Uncertainty in system design}
\label{secn:3.2.1}

Uncertainty in system design can be understood as the difference
between average and extreme system behavior.  System design must cope
with increasing uncertainty from two origins:

\begin{enumerate}
\item Uncertainty from the system's external environment exhibiting
  non-deterministic behavior, \eg time-varying load, dynamic change
  due to mobility and attacks.  How to figure out all possible
  security threats devised by an experienced hacker?
  
\item Uncertainty from the hardware execution platform which has
  inherently non-deterministic behavior owing to manufacturing errors
  or aging.  It also exhibits time non-determinism since execution
  times of even simple instructions cannot be precisely estimated due
  to the use of memory hierarchies and speculative execution.
\end{enumerate}

\begin{wrapfigure}[12]{r}{0.42\textwidth}
  \centering
  \includegraphics[width=0.4\textwidth]{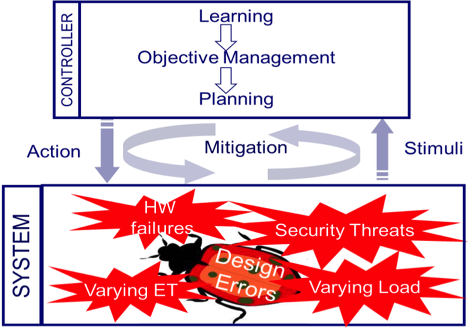}
  \caption{The principle of adaptive control}
  \label{fig:6}
\end{wrapfigure}

Uncertainty directly affects predictability, the degree to which
qualitative or quantitative system properties can be asserted (see the
discussion in \secn{4.1}).

Current state of the art in automated (and thus critical) system
design consists in making a detailed analysis of all the potentially
dangerous situations by clearly distinguishing between the ones the
system can cope with and confiding all the rest in human operators.
The analysis considers possible worst-case critical situations for
which the designer should implement corresponding mitigation
mechanisms and foresee the needed resources, \eg using redundancy.
This often results in over-provisioned, over-engineered systems, with
high production and operation costs.

For modern critical systems it is practically impossible to foresee at
design time all the possible hazards in system's lifetime due to poor
predictability.  Even if the sought degree of automation for a car is
the same as for an aircraft, the openness of car systems makes
impossible the static prediction of all potentially bad situations.
An additional reason for breaking with this design paradigm is that
the current divide with critical and best engineering effort is not
any more affordable both technically and economically: it is an
obstacle to increasing system integration \cite{ref6}.

\subsubsection{The principle of adaptive control}
\label{secn:3.2.2}

In the face of this situation, it is necessary to give up the
objective of guaranteeing trustworthiness at design time and seek a
better integration between critical and less critical features.  An
alternative, more realistic avenue comes from the concept of
adaptivity.  It consists in enforcing properties through the use of
intelligent controllers that continuously monitor the behavior of the
system and when a mishap is detected, they steer the system so as to
mitigate catastrophic effects.  Adaptive control has originated in
control theory \cite{ref23}.

An adaptive controller combines in a hierarchical manner three basic
functions (\fig{6}):

\begin{enumerate}
\item The central function is \emph{objective management} that
  consists in choosing for a given state the best objective by
  applying a multi-criteria optimization algorithm.  This algorithm is
  applied to a predictive model of the controlled system.

\item The \emph{planning function} is activated by the objective
  management in order to execute a mission for achieving an objective.

\item The \emph{learning function} is used to continuously update
  knowledge about the controlled system.  In particular, based on the
  monitored behavior it estimates parameter values of the objective
  manager.
\end{enumerate}

Adaptive control finds numerous applications in systems engineering to
enhance system trustworthiness and optimality.  For instance, to
ensure security, early warning mechanisms are used to learn and build
profiles of system users.  So, they can detect abnormal situations
caused by attacks or spyware and take adequate measures to mitigate
their effect \cite{ref24}.

We have applied adaptive control to systems integrating both critical
and best effort services.  The controller handles a provably
sufficient amount of global resources to: 1)~satisfy first and
foremost critical properties; and 2)~secondarily, to handle optimally
the available resources for best-effort services \cite{ref25,ref26}.

We believe that adaptivity is the technical answer to the demand for
both integrated mixed criticality systems and for trustworthiness
despite uncertain/unpredictable environments.


\section{Knowledge-Based Design}
\label{secn:4}

The idea is to design a system with adaptivity in mind by
incorporating the principle in the overall system architecture.  Not
only run-time knowledge, but also knowledge about the designed system
is applied for the detection of critical events and the enforcement of
properties.  Such an approach should be more effective than using
adaptive controllers external to the system.  Knowledge-based design
takes special care for accountability: at each design step we should
know which essential properties hold.  So, for the designed system we
know which properties are guaranteed at design time and which ones are
left to be monitored and possibly enforced at run time.

A system is deemed correct if knowledge both at design time and run
time about the system allows inferring satisfaction of its
requirements.


\subsection{The concept of knowledge}
\label{secn:4.1}

Knowledge is ``truthful'' information that can be used to
understand/predict a situation or to solve a problem.  Truthfulness
cannot always be asserted in a rigorous manner.  Nonetheless, we can
distinguish degrees between fully justifiable knowledge and empirical
knowledge.  Mathematical knowledge has definitely the highest degree
of truthfulness.  A theorem, \eg the Pythagorean theorem, is true
forever\mdash modulo acceptance of the axioms of Euclidian geometry.
Scientific knowledge is a generalization of experimental facts, \eg
Newton's laws, and as such it is falsifiable.  Then comes empirical
knowledge, which is not theoretically substantiated but is found to be
useful by experience.  Most common human knowledge is empirical, \eg
common sense knowledge, but also knowledge from machine learning can
be considered to a large extent as empirical.

Furthermore, knowledge may be declarative or procedural, regarding the
form it can take.  Declarative knowledge is a relation (property)
involving entities of a domain, whereas procedural knowledge describes
information transformation in a stepwise manner.  Typical examples of
declarative knowledge are the law of conservation of energy, a program
invariant or an architecture pattern.  Examples of procedural
knowledge are algorithms, design techniques, cooking recipes.


\subsection{Generated and applied knowledge in system design}
\label{secn:4.2}

\begin{figure}
  \centering
  \includegraphics[width=0.6\textwidth]{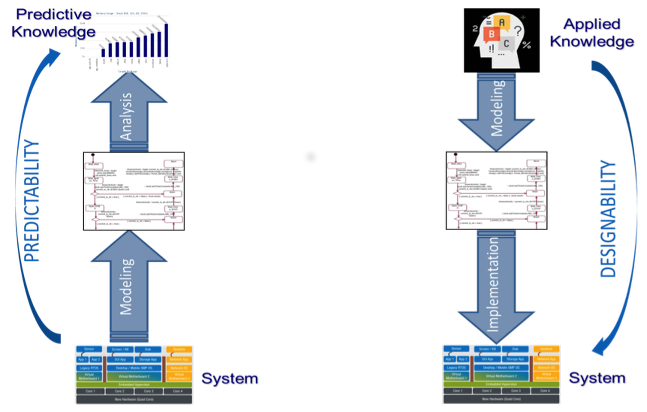}
  \caption{Bottom-up knowledge generation and the top-down knowledge
    application process}
  \label{fig:7}
\end{figure}

We discuss the principle of knowledge generation and application in
systems design (\fig{7}).

Designers study systems to generate knowledge about their behavior \eg
verification, performance evaluation.  The produced knowledge may be
invariants, performance evaluation measures, knowledge from learning
or statistical methods.

Rigorous knowledge generation from a system involves two steps.

\begin{enumerate}
\item The first step consists in modeling aspects of the system to be
  studied.  Models may be mathematical, \eg equations, or executable,
  \eg a piece of software.

\item The second step consists in analyzing the model to extract
  usable knowledge.  Analysis is often carried out using computers and
  can lead to either declarative or procedural knowledge.  For
  instance, declarative knowledge may be the fact that ``The door is
  always closed when the cabin moves'' or estimated latency.  An
  example of procedural knowledge is a testing scenario generated to
  validate a given system property.
\end{enumerate}

Consequently, our ability to predict system properties is limited by
two factors: 1)~the ability to model the studied system; 2)~the
ability to analyze models using computationally tractable methods.

Note that the terms \emph{descriptive} and \emph{predictive knowledge}
are often used to distinguish between the knowledge of the model and
the knowledge extracted by analysis.  Uncertainty is directly related
to our ability to build faithful models while predictability depends
on both the degree of uncertainty and the ability to extract usable
knowledge.

An important issue often overlooked by engineers, is to what extent we
can predict system properties and how their prediction is impacted by
modeling and computational limitations.

\begin{figure}
  \centering
  \includegraphics[width=0.6\textwidth]{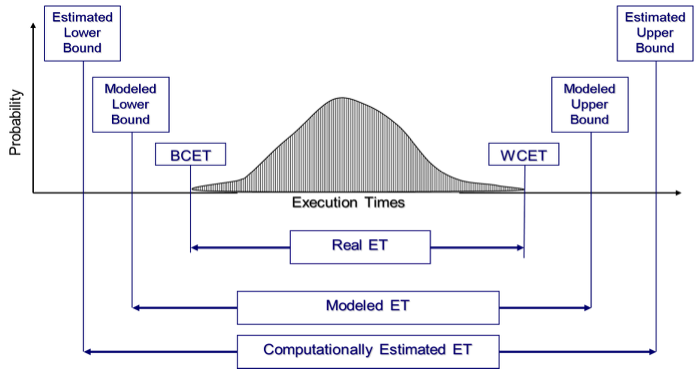}
  \caption{Real, modeled and computationally estimated execution times}
  \label{fig:8}
\end{figure}

Consider for example a central problem in critical systems engineering
that is the prediction of WCET (Worst Case Execution Times) and BCET
(Best Case Execution Times) for a software running on a given hardware
platform.  The rigorous process consists in building a model of the
mixed hardware/software system and then analyzing it to estimate
execution times \cite{ref27}.  As already explained, tractable models
can represent only some faithful abstraction of the real system.  This
means that execution times of a model are safe approximations of the
actual execution times.  Additionally, when analysis techniques are
applied, \eg by abstract interpretation, their effective application
requires further approximation.  So, as shown in \fig{8}, the
precision of the computed WCET and BCET depends on the precision of
both modeling and analysis.  It is important to note that the
application of such a rigorous approach to sophisticated architectures
may result in poor precision and practically useless results.

\begin{wrapfigure}[14]{r}{0.5\textwidth}
  \centering
  \includegraphics[width=0.5\textwidth]{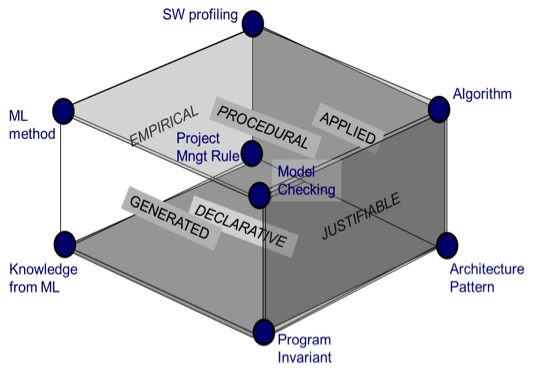}
  \caption{Types of knowledge used in system design}
  \label{fig:9}
\end{wrapfigure}

Often the knowledge generation process can be simplified when the
model is a conceptual one and/or the analysis is simply by reasoning.
For instance, by reading a program and analyzing assignments of
variables one can find that if some condition holds a variable takes a
certain value.  These ``superficial invariants'' prove to be useful
knowledge for proving other ``deeper'' properties.

Designers successively apply knowledge at each step of a design flow.
System design can be seen as a knowledge application/transformation
process starting from requirements and usable knowledge.  Knowledge is
progressively enriched with new knowledge generated by reasoning, by
construction or by analysis in order to obtain an implementable model.

As a rule, knowledge application in design involves two main steps.

\begin{enumerate}
\item The first step consists in building from some specification a
  system model meeting the requirements.  The main difficulty in this
  step may come from translating ambiguous specifications expressed in
  a natural language into a rigorous system model.

\item The second step involves computational complexity; it consists
  in extracting from the system model sufficient knowledge for an
  implementation meeting the requirements.
\end{enumerate}

Consequently, our ability to design systems is limited by two factors:
1)~the ability to build models meeting the requirements; 2)~the
ability to generate from these models, using computationally tractable
methods, sufficient knowledge for implementing a system.  Thus comes
the concept of \emph{designability}: to what extent is it possible to
rigorously build systems?

System engineers often overlook designability issues.  For complex
systems, requirements elicitation is probably the most critical design
step.  But even when we come up with adequate system models,
determining correct implementations may require significant analysis
effort as illustrated by the following example.

Consider the problem of finding trustworthy and optimal deployments of
some application software on a given multicore platform (\fig{2}).  As
explained in \secn{2.2}, a deployment is a mapping and an associated
scheduling algorithm.

Note that deployment trustworthiness and optimization can be assessed
only for known WCET.  However, WCET can be estimated only for known
deployments.  The WCET for a statement of a given task is the sum of
the WCET for execution in isolation and of the waiting time of the
task.  The latter depends on the deployment function due to resource
sharing with other tasks.  Thus, there is a cyclic interdependency
between deployments and WCET making the search for trustworthy and
optimal deployments an extremely hard problem for multicore platforms
\cite{ref28}.  This cyclic dependency can be broken for simple monolithic
systems, \eg flight control software for most Airbus types runs on bare
metal.

Note that the general knowledge application process may be simplified
when specifications are well understood or the implementation follows
a well-defined pattern.  This is the case when we apply a theorem, an
algorithm or we reuse a component in the design process.

\Fig{9} illustrates the three different kinds of knowledge presented
in this section.  Awareness that design is a knowledge
generation/application process should allow a more unified and
appropriate use of knowledge to achieve trustworthiness and
optimality.


\subsection{The Principle of Knowledge-Based Architecture}
\label{secn:4.3}

We define the two types of knowledge combined in a knowledge-based
architecture.

\begin{itemize}
\item \emph{Design-time knowledge} is essentially declarative
  knowledge about the properties of the designed system.  These
  properties may be established at some design step, using
  verification/analysis or by construction, \eg by reusing components
  or provably correct architecture patterns.  Accountability of the
  design flow allows knowing which system properties hold for the
  designed system

\item \emph{Run-time knowledge} is generated by monitoring system
  execution and deducing facts about the running system such as the
  violation of some property or knowledge generated by application of
  learning techniques.
\end{itemize}

\begin{figure}
  \centering
  \includegraphics[width=0.6\textwidth]{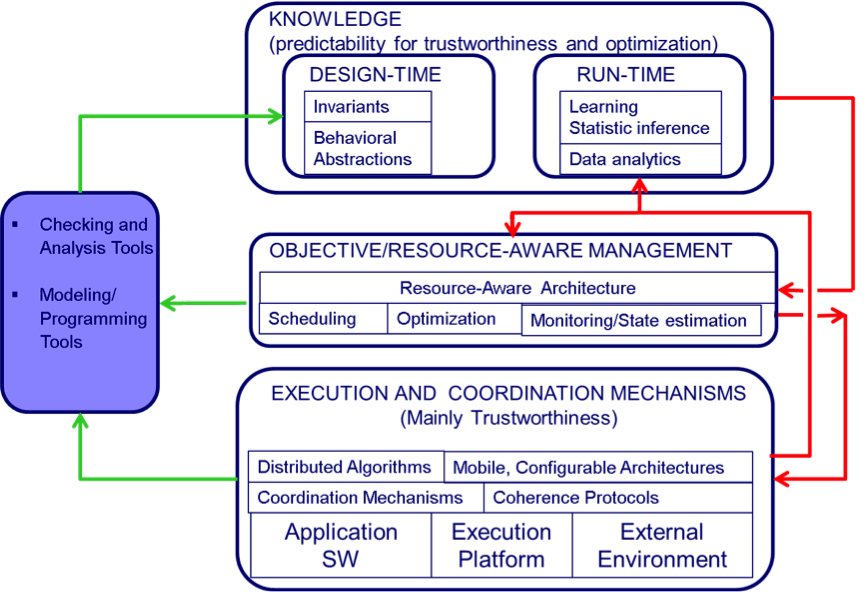}
  \caption{The principle of a knowledge-based logical architecture diagram}
  \label{fig:10}
\end{figure}

\Fig{10} depicts the principle of a knowledge-based logical
architecture diagram.  The decomposition into three layers is inspired
from adaptive architectures.  The upper layer is a repository for both
design-time and run-time knowledge.  It keeps updated the run-time
knowledge and combines the two types of knowledge to support the
management process of the second layer.

The middle layer includes execution and coordination mechanisms with
associated methods used by the application software to interact with
the platform and the external environment.  It is equipped with a
predictive system model.  It receives relevant knowledge from the
upper layer so as to manage objectives and resources.  Critical
objectives deal with meeting hard real-time constraints and coping
with critical failures and security threats.

The bottom layer integrates basic execution and coordination
mechanisms with associated methods.  It is important that their
functional correctness is established at design time.  This layer
receives orders mainly from the middle layer and sends back
information to both middle and upper layers (in red lines).

This schematic architecture leaves many issues open. 

Some issues have to do with striking the right balance between
design-time and run-time knowledge in order to achieve correctness.
It is preferable that predictable critical properties be established
at design time.  On the contrary, violation of properties involving a
high degree of uncertainty should be detected on-line and mitigated.
Of course, other criteria may influence this balance, such as cost
effectiveness and the sought degree of availability.  Runtime
enforcement of a property may require interruption of service, \eg
using a fail-safe mechanism.

\begin{figure}
  \hfill
  \includegraphics[width=0.45\textwidth]{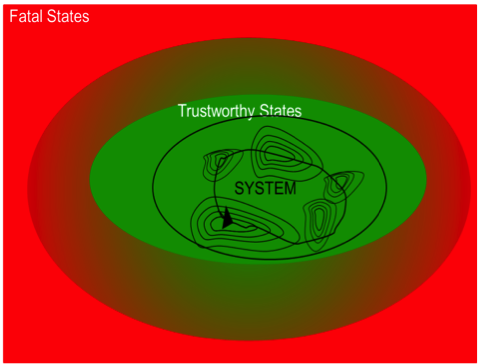}
  \hfill
  \includegraphics[width=0.45\textwidth]{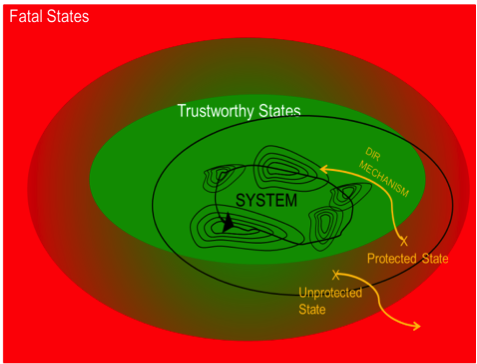}
  \hspace{\fill}
  \caption{Correctness at design time vs. partial correctness}
  \label{fig:11}
\end{figure}

\Fig{11} illustrates two extreme cases: correctness at design time and
partial correctness.  The system state space is partitioned between
trustworthy states (green region) and non-trustworthy states.  The
latter can be either 1)~fatal states, \ie states at which system
trustworthiness is definitely compromised, or 2)~non-fatal states that
can be detected early enough so that a timely recovery into a
trustworthy state is enforced at run-time using some DIR (Detection,
Isolation and Recovery) mechanism.  Correctness at design time
corresponds to the ideal case where all the possible system states are
proven to be trustworthy.  In practice, avoidance of fatal states is
achieved thanks to enhanced predictability in order to detect as early
as possible flawed states and determine isolation and recovery
strategies.  A system is deemed correct if it always remains in
trustworthy or non fatal states from which timely recovery is assured.

Other architectural issues have to do with performance and, in
particular, the ability to meet hard real-time constraints.
Determining which objective to choose in order to cope with a critical
situation and planning the associated mission, may require non
negligible computational power and time.  It is important to find
trade-offs between response times and precision of the management and
planning processes.  A crude and fast response is often better than a
refined and slow one.

Finally, practical issues will ultimately weigh the relevance of the
approach.  These include the implementation of knowledge-based
architectures and the integration of the two types of knowledge in a
platform.  For selected application areas, we need scalable knowledge
management techniques allowing enhanced predictability as well as
online control methods with sufficiently low overhead footprint.


\section{Discussion}
\label{secn:5}

We have amply explained why future autonomous systems cannot be
designed as classical critical systems.  Current trends require novel
rigorous design methodologies for open autonomous interconnected
systems involving embedded supercomputers, AI algorithms and receiving
data from the Cloud.  We also need new trustworthiness assessment
techniques and standards for third party certification.

Stringent predictability requirements of safety standards such as
ISO~26262 and DO~178B preclude their application to IoT autonomous
systems.  Although they cannot guarantee absence of bugs, these
standards allow checking the quality of the development process and
provide model-based guarantees that the system can cope with
predictable mishaps compromising its safety.  Clearly, they cannot
handle machine learning software as it cannot be checked against
requirements.

We currently lack standards for autonomous systems.  Although
certification by independent labs is mandatory, even for home
appliances like toasters, the automotive and medical device industry
are exempted from third-party certification.  Robocars are
self-certified by their manufacturers following guidelines and
recommendations issued by authorities.  Some autonomous car
manufacturers consider that safety can be guaranteed only through
testing an extremely large number of scenarios.
  
Interestingly enough, a recent article by Mobileye \cite{ref29}
advocates model-based safety as the only realistic approach for
validating autonomous vehicles.  Furthermore, it adheres to the
well-understood position that testing-based approaches will require
exorbitant time and money to achieve sufficient evidence of
reliability \cite{ref30}.  Although the article only very partially
tackles the multitude of issues raised by model-based design, it has
the merit of posing the problem of rigorous safety evaluation and has
already initiated controversial discussion in the media, \eg
\cite{ref31}.  It is a pity that the current debate focuses on the
reliability of learning techniques and associated sensory devices
while it completely overlooks key system design issues pertaining to
global trustworthiness.

We believe that there is a risk that under the market and business
pressure, the competent authorities accept the generalized deployment
and use of self-certified autonomous systems without any conclusive
evidence about their trustworthiness.  A strong argument in favor of
this can be that fully autonomous systems may be statistically safer
than semi-autonomous systems (\emph{``In the distant future, I think
  people may outlaw driving cars because it's too dangerous''}\mdash
  Elon Musk, May 18, 2015).

The advent of IoT is a great opportunity to reinvigorate Computing by
focusing on autonomous system design.  This certainly raises
technology questions but, more importantly, it requires building new
foundation that will systematically integrate the innovative results
needed to face increasing environment and mission complexity.  A key
idea is to compensate the lack of human intervention by adaptive
control.  This is instrumental for system resilience: it allows both
coping with uncertainty and managing mixed criticality services.  Our
proposal for knowledge-based design seeks a compromise: preserving
rigorousness despite the fact that essential properties cannot be
guaranteed at design time.  It makes knowledge generation and
application a primary concern and aims to fully and seamlessly
incorporate the adaptive control paradigm in system architecture.


\section*{Aknowledgement}

The author thanks Dr Simon Bliudze for his valuable comments and
suggestions for improving the paper.

\bibliographystyle{eptcs}
\bibliography{sifakis-metrid2018}
\end{document}